\documentclass[sw]{agu2001}

\usepackage{amssymb}
\usepackage{graphicx}
\usepackage{epstopdf}

\newcommand{\figheight}{3.5in}

\authorrunninghead{SMITH AND SCALO}
\titlerunninghead{EVA X-ray Flare Hazard}

\authoraddr{D.~S.~Smith, Lunar and Planetary Laboratory, University of
Arizona, 1629 East University Blvd, Tucson, AZ 85721 (dss@lpl.arizona.edu)}

\authoraddr{J.~M.~Scalo, Department of Astronomy, The University
of Texas at Austin, 1 University Station C1400, Austin, TX 78712
(parrot@astro.as.utexas.edu)}

\begin{document}


\title{Risks due to X-ray Flares during Astronaut Extravehicular Activity}

\author{David S.~Smith\altaffilmark{1}}
\affil{Department of Astronomy, The University of Texas at Austin, Austin,
Texas 78712}

\altaffiltext{1}{Present affiliation: Lunar and Planetary Laboratory,
University of Arizona, Tucson, Arizona 85721}

\author{John M.~Scalo}
\affil{Department of Astronomy, The University of Texas at Austin, Austin,
Texas 78712}

\begin{abstract}


Solar hard X-ray flares can expose astronauts on lunar and deep space
extravehicular activities (EVAs) to dangerous acute biological doses. We
combine calculations of radiative transfer through shielding materials
with subsequent transfer through tissue to show that hazardous doses,
taken as $\ge 0.1$ Gy, should occur with a probability of about 10\%
per 100 hours of accumulated EVA inside current spacesuits.  The rapid
onset and short duration of X-ray flares and the lack of viable precursor
events require strategies for quick retreat, in contrast to solar proton
events, which usually take hours to deliver significant fluence and can
often be anticipated by flares or other light-speed precursors. Our
results contrast with the view that only particle radiation poses
dangers for human space exploration.  Heavy-element shields provide
the most efficient protection from X-ray flares, since X-rays produce
no significant secondary radiation.  We calculate doses due to X-ray
flares behind aluminum shields and estimate the required shield masses
to accompany EVA rovers.

\end{abstract}

\begin{article} 

\section{Introduction}

The risk for space travelers due to solar radiation, as well as Galactic
cosmic rays, has been studied extensively for nearly four decades, but the
predictability and frequency of potentially lethal doses of high-energy
solar radiation is still far from understood. A long series of studies
suggest that the most energetic solar proton events (SPEs) \citep{reedy96,
shea+smart90, shea+smart00, miroshnichenko03, reames99} can produce lethal
biological doses that require careful consideration of radiation risks
for manned space travel \citep[e.g.][]{silberberg+tsao79, letaw+89,
dyer+96, schimmerling+96, badhwar97, cucinotta+01, miroshnichenko03,
cucinotta+04, deangelis+04a, deangelis+04b, getselev+04, wilson+04,
cougnet+04, johnson+05, obrien05, kim+06}.

The August 1972 SPE, often taken as the standard high-fluence event
for protection studies, was not an isolated anomaly. The February 1956,
November 1960, October 1989, and October-November 2003 events produced
solar-particle fluxes sufficiently large that an astronaut on the moon
protected by only a spacesuit would likely have perished, and many events
approaching these fluences have been recorded \citep{shea+smart00,
miroshnichenko03}.  The durations of these events (many hours, see
\citealt{miroshnichenko03}) are large enough that evacuation to shelter
would prevent serious exposure to an astronaut on an EVA. Observations
of associated lightspeed precursors (photon flares and CME eruptions),
while not always available for the strongest SPEs \citep{reames99},
could allow further time for escape.

Given the number of studies that have taken on the computational
difficulties of the propagation of solar cosmic rays in the inner
heliosphere and the resulting biological doses, it is surprising that the
simpler study of the effects of ionizing photons, in particular X-rays
from solar flares \citep{haisch+91, hudson91}, have not been considered.
It is easy to show that the fluence from the most energetic hard X-ray
flares of the last 50 years would result in an amount of energy absorbed
per gram of tissue larger than the lethal dose for an unprotected human
at 1 AU.  Acute doses for various biological endpoints are well studied,
especially because of their relevance to oncological research. The
frequency of very energetic flares is similar to that of the most
energetic SPEs; their durations are so small that evacuation is much
more problematic; and there are no early and reliable precursors that
can be used to predict their onset as in the case of SPEs. Furthermore,
the shielding requirements are nontrivial and somewhat different from
those for solar energetic particles, preferring material of
high atomic number, in contrast to the mostly polymer construction of
current spacesuit designs. Bone surrounding blood-forming marrow might
mitigate hematological effects, but the risk of carcinogenesis remains
significant. We calculate here the expected doses from energetic solar
X-ray flares and use their observed frequency-fluence relation to estimate
the probability per unit time of a hazardous flare exposure.

\section{The Largest Solar Flares}

Solar flare photons span energies between 10$^{-1}$ and 10$^6$ keV
\citep{haisch+91, hudson91, kanbach+93, ryan00}.  RHESSI observations
\citep[e.g.][]{battaglia+05} show that the spectrum of hard X-rays
(HXRs) from solar flares extends to energies as low as 10--25 keV,
where ``hard'' refers to the power-law part of the spectrum produced by
particle acceleration rather than thermal processes.  The HXR spectrum
is usually fit by the form $F(E) \sim E^{-p}$ \citep{crosby+93a, lee+93,
bromund+95, veronig+02, qiu+04}. The range of the estimated power-law
index (log-log slope) $p$ is large for HXR flare spectra, with $2 < p <
6$ and a median around 3.5. Additionally, a small correlation between $p$
and total X-ray output has been suggested by RHESSI \citep{battaglia+05}.

X-class solar flares (the most energetic flares) have typical durations of
about 5--30 min \citep{crosby+93a, veronig+02}, with the more energetic
flares tending to last longer. As with spectral slope, flare durations
vary widely, but even with the longest of these durations the dose
should be considered acute (i.e. effects have rapid onset), in contrast to
low-level, extended exposure to Galactic cosmic rays. The acute lethal
dose is almost always independent of dose rate \citep{sparrow+67},
so we assume that the nonlethal acute dose depends on total fluence,
not flux, and hence on the flare total energy release.

The total photon energy release $W$ in flares is difficult to estimate
and varies by at least a factor of 10$^8$, but a large number of studies
using EUV, soft X-ray (SXR), and HXR satellite events roughly agree
that the differential distribution $dN/dW$ of flare energy releases is a
power law with index about $-1.6$ to $-1.8$ over at least six orders of
magnitude in total energy release $W$ \citep{hudson91, lee+93,
crosby+93a,
bromund+95, aschwanden+00, lin+01, guedel+03, qiu+04}.

The largest X-ray releases observed among contemporary solar flares is
$\sim 10^{32}$ erg \citep{hudson91, crosby+93b}.  Eleven X-class flares
occurred during the extraordinary solar outbursts between 18 October
2003 and 5 November 2003 \citep{gopalswamy+05}, with SXR releases in
the GOES 1--8 \r{A} (2--10 keV) band peaking at $2 \times 10^{31}$ erg for
the largest bursts.  Observations using the SORCE instrument's Total
Irradiance Monitor yielded a total flare energy at all wavelengths for the
28 October flare of $4.6\times 10^{32}$ erg \citep{woods+04}.  Radiation and
charged particles from these flares compressed the Earth's Van Allen
belt to within 20,000 km of the surface \citep{baker+04}, damaged the
orbiting Mars Odyssey communication instruments, and reduced polar ozone
levels significantly \citep{randall+05}.

Although most of our calculations are for a 10$^{31}$ erg flare, it seems
likely that much more energetic flares than considered here have occurred.
Such flares have been serendipitously discovered in other solar-like stars
with otherwise normal characteristics \citep{schaefer+00}. Upper
limits on proton fluences inferred from cosmogenic isotopes in lunar
samples \citep{reedy+83, reedy96}, tree ring records of $^{14}$C
\citep{lingenfelter+hudson80}, and the statistics of impulsive nitrate
events \citep{mccracken+01} suggest that the frequency-fluence relation
steepens for high-energy particle fluences above about 10$^{10}$
cm$^{-2}$, probably due to streaming-limited fluxes associated with
self-confinement by ion-wave interactions \citep{reames99}. In contrast,
no such limit, empirical or theoretical, has been established for photon
flares, other than an upper limit to avoid divergence of total energy
\citep{hudson91, aschwanden99}.

We adopt a mean recurrence time for 10$^{32}$ erg flares of 10 yr,
agreeing with the estimate of \citet{hudson91} and broadly consistent
with the dozen or so 10$^{31}$--10$^{32}$ erg events that have been
observed since GOES soft X-ray monitoring began in 1976.  The average HXR
frequency-energy release statistic $dN/dW$ is taken as a power law in $W$,
with log-log slope $-1.7$.  Thus, we take the mean time between events of
energy release $W_{31}$ (in units of 10$^{31}$ erg) as \begin{equation}
\tau(W) = 0.2\, W_{31}^{1.7}\ \mathrm{yr}.\end{equation}

\section{Methods}

For our calculations, solar flare photon number spectra are assumed
to be distributed as power laws, $E^{-p}$, with $2 < p < 6$ (see above). The
flare spectrum is assumed to extend from 10 keV to 511 keV.  The 10 keV
lower limit is taken to simulate the HXR flares, which often begin to
flatten to a thermal form around this energy, while the upper limit is
set high enough that a negligible number of incident photons are at higher
energies, even for the shallowest spectra (lowest $p$).  Henceforth, total
energy release quantities refer only to the energy between 10 and 511 keV.

We transport the incident ionizing radiation using a single-scattering
approximation.  The primary interaction processes at X-ray energies are
photoelectric absorption and Compton scattering.  The photoabsorption
opacity is much larger, so the radiative transfer can be computed by
employing an effective opacity, in which both Compton scattering and
photoabsorption are treated as absorption processes.  By including
the Compton cross section in the opacity, we in effect assume that the
material is optically thin to Compton scattering, such that at most a
photon will scatter no more than once before exiting.  Errors incurred
by this approximation are negligible, as we have verified using
Monte Carlo simulations \citep{smith+04}.  The Compton-scattering
coefficient was computed exactly using the Klein-Nishina formula,
while the photoabsorption cross section was approximated using the
empirical form of \citet{setlow+pollard62}:  \begin{equation}\sigma_p(E)
= 2.4 \times 10^{-30} (1 + 0.008 Z) \left(\frac{Z}{E}\right)^3\,
\mathrm{cm}^2,\end{equation} where $Z$ is the atomic number of the target
nucleus and $E$ is the energy of the incident photon in units of 511 keV.
We found this form to be an excellent fit to cross section data for light
elements and for energies higher than the K edge.  The K edges of carbon
and aluminum are at (respectively) 284 and 1560 eV \citep{henke+93} and
are well below our spectral cutoff energy of 10 keV, so our calculation
is in the energy regime that is consistent with the cross section fit.
The effective opacity of the material is then \begin{equation}\kappa(E)
= \frac{\sigma_p(E) + Z \sigma_c(E)}{A m_H},\end{equation} where $A$ is
the mean atomic mass in amu of the target material, $m_H$ is the mass of
a hydrogen atom, $Z$ is its atomic number, and $\sigma_c$ is the Compton
cross section.  The single-scattering approximation results in exponential
attenuation of the incident fluence.  In terms of the effective opacity
$\kappa$ and the areal density $\Sigma$ (g cm$^{-2}$) of the shielding
material, the attenuated energy fluence $F$ is \begin{equation}
F(E,\Sigma) = E \frac{dN}{dE} \exp[ -\kappa(E) \Sigma ].\end{equation}

After transport through the shielding material, the photon fluence is
converted to a biological dose by assuming that the remaining radiation
is absorbed by pure water.  Much of the damage by ionizing radiation
is thought to be ``indirect,'' involving chemical reactions initiated
by energy deposited in the bulk cell water or first hydration layer,
rather than ``direct'' ionization of DNA \citep{vonsonntag87, ward99},
although this terminology is now recognized as an oversimplification
\citep{fielden+oneill91}.  We then estimate the dose as a function of
areal density of the shielding by integrating the attenuated incident
photon energy spectrum over the effective opacity of water $\kappa_w(E)$
(calculated using the single-scattering method above): \begin{equation}
D(\Sigma) = \int_{10}^{511} \kappa_w(E)\; F(E,\Sigma)\; dE, \end{equation}
where the attenuated fluence $F$ is given above, and the limits of
integration are in keV.  This is essentially the skin dose.

\section{Minimum Hazardous Dose}

To estimate risks to humans exposed to solar flare X-rays, critical
levels of acute radiation exposure for various types of health outcomes
(e.g. hematologic damage, organ failure, cancer, and lethality)
are needed. Particularly useful is the minimum dose at which a given
enhancement in occurrence of a disease relative to the average population
occurs.  We use dosimetry units for which 1 gray = 100 rad = 10$^4$ erg
g$^{-1}$ absorbed.  For reference, a chest X-ray delivers around 0.1 mGy.
Since X-ray photons have small linear energy transfer, $dE/dx$, their
``quality factor'' or ``biological effectiveness'' is very close to
unity, independent of energy, so dose in Gy is approximately the dose
in sievert. The duration of exposures to flare X-rays of large fluence
will be relatively short (10--60 min in most cases), so we restrict the
discussion to evidence concerning acute  human radiation-syndrome data.

At doses above about 0.5 Gy, summaries of a number of sources of
data relevant to acute radiation syndrome in humans are available
\citep{alpen98, turner95}, as well as detailed specialized studies
\citep{dickinson+parker02, satoh+96, dubrova+97, dubrova03}. There is
general agreement that severe damage, primarily hematologic and without
assured recovery, occurs around 1 Gy. Estimates of the whole-body acute
lethal dose vary from 2 to 5 Gy \citep{unscear01, alpen98, turner95}.
An upper limit to the radiation risk dose in an exposed individual is
the ratio of the average spontaneous rate of mutations over a large
number of genes to the induced mutation rate per Gy for low-LET (linear
energy transfer) irradiation, recognizing that the mutation rate is
highly variable among loci; this ratio is the mutation doubling dose.
An estimate using 135 human genes for spontaneous rates and 35 mouse genes
for induced rates \citep{sankaranarayanan+chakraborty00} gives a doubling
dose of 0.8 Gy. This dose is for chronic (i.e. continual) irradiation,
and a dose rate reduction of a factor of three is usually assumed for
acute doses, suggesting a doubling dose of 0.3 Gy for acute irradiation.
However this estimate remains uncertain and subject to various definitions
\citep{sankaranarayanan+chakraborty00, unscear01}.

Most work on X-ray and gamma-ray radiation risk to exposed
individuals (not genetic disease endpoints) comes from studies of
carcinogenesis. There is little doubt that the incidence of cancer
due to radiation-induced genomic instability rises with acute
dose above 0.2--0.3 Gy.  Below this dose there is continued debate
whether there is a ``linear-no-threshold'' relation between risk and
ionizing-radiation dose, or whether doses below about 0.1 Gy result in
``adaptive response'' causing endogenous DNA damage prevention and
immune stimulation \citep{feinendegen05}. Alternatively, it is also
likely that risk at low doses is larger than the linear-no-threshold
extrapolation, even increasing with decreasing dose, because of
bystander effects, as reviewed in \citet{hall04}.  The situation is
further complicated by the existence of a significant fraction of
humans with predisposing mutations to cancers induced by ionizing
radiation \citep[e.g.][]{sankaranarayanan+chakraborty00}.  It is
certain that mutations, often at significant loci \citep{sparrow+72,
sankaranarayanan82, forster+02}, chromosomal abnormalities in
blood lymphocytes \citep{violot+05}, and clustered DNA damage
\citep{sutherland+00} occur at much smaller doses in the range 0.01
to 0.1 Gy, but their impact on health risk has been difficult to
assess because of the evidence for adaptive response at low doses. A
compelling discussion by \citet{brenner+03} argues that there is good
human epidemiological evidence for increased cancer risk at an acute
X-ray or gamma-ray dose of 0.05 Gy, and reasonable evidence for enhanced
risk above 0.01 Gy, although the risk enhancement is in the 1--10\% range.

Acute critical doses for significantly increased risk for other
disease endpoints may also be of order 0.1 Gy or less, especially for
hematological diseases. For example, the summary of delayed somatic (i.e.
bodily) effects due to acute doses by \citet{hanslmeier02} indicates that
gastrointestinal tract syndrome (leading to loss of digestion ability,
bleeding ulcers, and diarrhea) sets in at 0.1 Gy for X-rays.

The risk of a particular \emph{genetic}-disease endpoint per
Gy of irradiation is more difficult to estimate, since the
disease-specific induced mutation rate varies greatly and most
models assume mutation-selection equilibrium.  Inspection of risk
estimates in humans and animal models for a number of genetic-disease
classes that include 26 human disorders encompassing 135 genes
\citep{sankaranarayanan+chakraborty00} indicates that risks for
genetic-disease endpoints at 0.1 Gy due to acute X-ray doses are
probably much smaller than the risks for somatic disease in an exposed
individual discussed earlier.  These estimates include corrections for
potential recoverability and concentrate on low-LET radiation, X-rays and
$\gamma$-rays.  The results are given for chronic irradiation, even though
they are based on experiments with high dose-rate irradiation.

The ``dose-rate reduction factor'' for acute doses is believed to
be roughly a factor of three, a factor which is well established in
studies of specific-locus mutations.  Using this factor to correct the
results, it is found that risks in terms of the excess over an average
population for acute doses are typically 1\% per Gy.  But these results
were typically obtained at equivalent acute doses in the range 0.5--3 Gy,
so linearity of the above risk with dose cannot be assumed.

A review of evidence for enhanced mutation rates in human populations
exposed to doses as low as 0.25 Gy in the Chernobyl accident and nuclear
weapons tests in Kazakhstan is given in \citet{dubrova03}.  But the main
point is that the risks for genetic disease endpoints at 0.1 Gy might
be smaller than the risk for somatic disease in an exposed individual.

Given that the linear increase of cancer risk with dose for acute doses
above 0.1 Gy seems unequivocal  and that the threshold for delayed
somatic-disease endpoints appears to also be about 0.1 Gy, we adopt
this biological dose as an upper limit for significant risk increase
due to X-rays, with the understanding that enhanced risk for several
disease endpoints, but especially cancer \citep{brenner+03}, may
still be significant at lower doses.  An important unresolved point is
that adoption of a critical dose depends on the risk enhancement that
one is willing to accept; our reading of the literature suggests a 10\%
enhancement in potentially fatal disease endpoints at 0.1 Gy, and less
than 1\% enhancement for genetic disease endpoints at this dose.

\section{Results}

\subsection{Risk Estimate}

\begin{figure} \centering
\includegraphics[height=\figheight,angle=270]{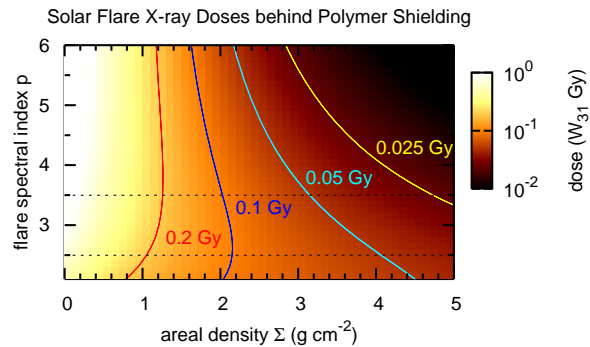} \caption{Acute
biological doses behind polymer shielding (representative of current
spacesuits) due to a 10$^{31}$ erg X-ray flare as a function of flare
spectral index, $p$, and areal density, $\Sigma$.  The dose is roughly
independent of $p$ for shields with areal densities smaller than about 2
g cm$^{-2}$.  For larger shielding columns, the dose becomes sensitive to
the spectral shape because more of the incident spectrum is attenuated,
and hence reshaped, before being absorbed by the model water column.
Areal densities of polymer in excess of 2 g cm$^{-2}$ are needed to
reduce the X-ray dose to below our adopted maximum acceptable acute dose
of 0.1 Gy. } \label{fig:eva1} \end{figure}

We can now compare our adopted upper dose limit of 0.1 Gy with that
received behind polymer shielding, which is representative of current
spacesuit design, such as the space shuttle Extravehicular Mobility Unit
\citep{ross+97}. We use pure carbon as a proxy for the mostly polymer
construction of spacesuits because only the carbon atoms in polymers
significantly absorb X-rays.  The results, assuming a 10$^{31}$ erg
solar flare, are shown in Fig.~\ref{fig:eva1}.

The typical areal densities of spacesuit components ($\sim$ 0.5--1.5
g cm$^{-2}$) provide little protection during a large solar flare. A
relatively common 10$^{31}$ erg flare would deliver over 0.2 Gy behind the
current spacesuit---twice our adopted upper limit.  A thicker spacesuit
could reduce this dose to 0.1 Gy, but larger flares do occur, albeit less
often. Bolstering spacesuits simply decreases the frequency of dangerous
doses and does not eliminate the threat.

What is the likelihood of being exposed to doses above 0.1 Gy? If
we use for simplicity a typical flare spectral index of 3.5, we find
that the approximate dose behind polymer shielding of areal density
$\Sigma \gtrsim 0.3$ g cm$^{-2}$ is \begin{equation} D(\Sigma,W) =
0.21\, W_{31} \Sigma^{-1.4}\ \mathrm{Gy},\end{equation} where $W_{31}$
is the flare energy release in units of 10$^{31}$ erg, and $\Sigma$ is
the areal density of the shielding material in g cm$^{-2}$.  Using the
flare energy-recurrence frequency relation given in the discussion of
flare properties to eliminate $W_{31}$, the mean time between flares
delivering \emph{at least} a given dose $D_{0.1}$ behind an polymer
shield of areal density $\Sigma$ is \begin{equation} \tau(\ge D) = 0.08\,
D_{0.1}^{0.7}\, \Sigma\ \mathrm{yr},\end{equation}  where $D_{0.1}$ is
the dose in units of 0.1 Gy. If we then assume that time between flare
events is typically much larger than the duration of exposure and that
flares occur at random, the probability of exposure to a dose of at least
the critical dose is \begin{equation} P(D,\Sigma) = 1 \times 10^{-3}
D_{0.1}^{-0.7} \Sigma^{-1}\end{equation} per hour of EVA.

So in just 100 hours of EVA in the current spacesuit, an astronaut would
accumulate a 10\% risk of a dangerous exposure to solar flare X-rays.

\subsection{X-rays Compared to Solar Energetic Particles}

The previous section discusses only the risk due to photons from
flares. The risk due to solar energetic particle events (SPEs)
is certainly not negligible, but we argue that the risk during EVAs is
significantly smaller than estimated here for hard X-rays from flares.
The largest SPEs, such as the 14 July 2000, Feb 1956, and Aug 1972
events, had 1 AU fluences in the range 10$^{10}$--10$^{11}$ cm$^{-2}$
\citep{miroshnichenko03}, although per-particle energies were 100 times
smaller than for Galactic cosmic rays.  \citet{silberberg+tsao79}
estimated the incidence of flares that produce SPEs with 1 AU doses
greater than 1 Gy as about one per decade, similar to what we estimate
for 10$^{32}$ erg photon flares that produce 2 Gy doses, as shown
in the Fig. 1.  A similar frequency for SPEs with fluences above
10$^{10}$ cm$^{-2}$ can be derived from the recorded number of large
events \citep{reedy96, shea+smart90, miroshnichenko03}, nitrate ice
core reconstruction covering several centuries \citep{mccracken+01},
and probability models \citep{feynman+93, feynman+02}.  In contrast,
10$^{31}$ erg X-ray flares that also require substantial shielding are
about 50 times more frequent.

Spacesuits are the last line of defense until shelter is reached, so
exposures during a flare will depend on the ratio of the time to reach
shelter to the time to deliver the total fluence.  Figure \ref{fig:eva2}
shows that the doses behind aluminum shielding are significantly smaller
than that received inside spacesuits of the same areal density because
of the strong dependence of the photoabsorption cross section on atomic
number of the target material.  But X-ray flares leave only 10--30 min
to reach shelter before the total fluence is delivered.

Hard X-ray flares are impulsive, with rise times of minutes
or less.  The time to withdraw to adequately shielded shelter is
very small.  A reliable energetic hard X-ray flare precursor signature
occurring more than an hour before the flare maximum would be needed
for this purpose. There are many signatures that have been proposed
as flare precursors \citep{martin80, simnett93}, such as
UV brightening, soft X-ray enhancements, microwave radio signatures,
H$\alpha$ filament disturbances, strong magnetic shear, sunspot motions,
and the beginnings of chromospheric mass ejections (CMEs), whose onset is
now known to slightly precede an associated flare on average.  Most of
these precursors are only observable for less than a few minutes before
the onset of the flare, so do not give sufficient warning and would
require elaborate monitoring systems.  Some radio precursors, such as
polarization signatures, are observed up to tens of minutes before a
flare, but are not observed in the majority of flares.  Similarly, for
all precursor signatures there are flares seen without the precursor
and observations of ``precursors'' that are not followed by a flare;
no precursors are necessary and sufficient \citep{golub+pasachoff97}.

Even more seriously, none of these precursors are predictors of the
energy release of the flare itself, so given the large frequency of
flares that pose no biological hazard, use of these precursors would
likely result in a large false-alarm rate.

Most probabilistic approaches for flare prediction are based on
a combination of historical rate of flaring for a given sunspot
classification group and additional information such as shear, magnetic
topology, and previous large-flare activity \citep{gallagher+02}.
A more recent Bayesian method relies only on flare event statistics
\citep{wheatland05}.  These methods are most suited for probabilistic
prediction of quantities like the number of flares of a certain class
in a given year, but not for EVA hazard warnings.  For example, the
Bayesian method would have predicted a 20\% probability for an X-class
flare on 4 Nov 2003, using data up to one day before, including a
highly clustered series of strong and weak flares in the week before,
yet actually the most energetic flare in several decades was about to
occur \citep{wheatland05}, a flare almost an order of magnitude more
energetic than the model flare used in the calculations reported here.

Such prediction algorithms might be useful for policies requiring no EVAs
in windows of a week or so, when the probability of a large flare can
be somewhat more accurately estimated, but this could greatly curtail
manned exploration, depending on where the threshold for significance
is placed and the reliability of the prediction. For example, less
reliable predictions would require lower thresholds for significance
(and thus higher alarm rates) to maintain acceptable risk levels.

Because of diffusive propagation through the heliosphere, the first
particles from SPEs arrive at 1 AU more than an hour after the initiating
event (when one is observed) and take hours to reach hazardous fluences
\citep[][Fig. 2.7, 12.9]{miroshnichenko03} \citep{reames+97}.  This allows
a simple and effective retreat strategy based on flare, CME, or even
sunspot precursors (although not all energetic SPEs are associated with
flares; \citep{reames99}).  Consequently, X-ray flares are more dangerous
during EVAs.

\subsection{Emergency Shielding}

\begin{figure} \centering
\includegraphics[height=\figheight,angle=270]{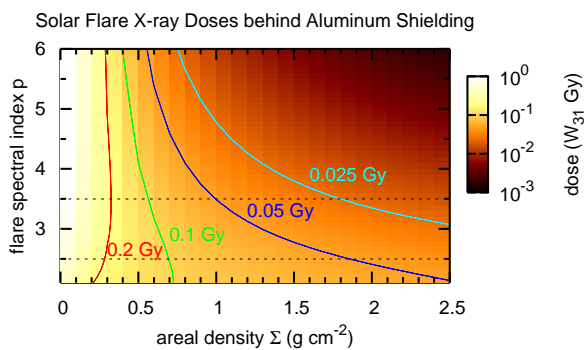} \caption{
Acute biological doses behind aluminum shielding (representative of
current possible radiation shelters) due to a 10$^{31}$ erg X-ray flare
as a function of flare spectral index, $p$, and areal density, $\Sigma$.
Aluminum has a higher atomic number than carbon (13 vs. 6) and absorbs
X-rays much more efficiently.  Only 2 g cm$^{-2}$ of aluminum shielding
would be required to reduce the dose to below 0.05 Gy for a 10$^{31}$ erg
flare.} \label{fig:eva2} \end{figure}

One simple protection solution would be to include a 2--3 m$^2$
shield of high-$Z$ material, such as aluminum, in EVA rover designs.
Heavy elements more efficiently stop X-rays than light elements, and no
significant secondary radiation is produced by X-rays, so we believe a
material such as aluminum to be optimal for this purpose.  Using the
results in Fig.~\ref{fig:eva2}, an aluminum shield thick enough to protect
against a 10$^{31}$ erg flare would have an areal density of at least
7 kg per square meter of shielding; for a 10$^{32}$ erg flare the mass
requirement would be 70 kg m$^{-2}$. Additionally the shield would have to
be articulated or detachable, in order to provide protection regardless
of the sun's position in the sky.  This shield would be employed during
evacuation, since it would be useless or even dangerous if a large SPE
followed the flare, since the high-$Z$ composition would enhance the
radiation dose from secondary particle production within the shield.

\section{Summary and Conclusions}

The risk to astronauts due to solar flare X-rays was before now mostly
unknown, in contrast to the highly studied effects of particle radiation,
such as solar protons and Galactic cosmic rays.  Here we have calculated
the transport of ionizing radiation through spacesuit material and present
acute biological doses due to X-rays that an astronaut could receive if a
large solar flare occurs during an EVA.  Based on the studies of somatic
and genetic disease risks due to acute doses of ionizing radiation, we
adopt a minimum hazardous dose of X-rays of 0.1 Gy and find that the risk
of receiving at least 0.1 Gy from an X-ray flare is roughly 10\% per 100
hours of accumulated EVA.  The onset and duration of X-ray flares is rapid
enough and possible precursors are unreliable enough that avoidance would
be difficult.  The simplest solution for X-ray protection on rover-based
EVAs could be the inclusion of a mobile body shield to supplement the
shielding provided by the spacesuit until shelter can be reached.

\begin{acknowledgements}

DSS was supported by the NSF Graduate Student Research Fellowship and
Harrington Doctoral Fellowship Programs. JMS was supported by the NASA
Exobiology Program, Grant NNG04GK43G.  This work was carried out as part
of the research of the NASA Astrobiology Institute Virtual Planetary
Laboratory Lead Team, which is supported through the NASA Astrobiology
Institute.

\end{acknowledgements}

\end{article} 


\end{document}